# 1. Introduction

Polymers are ubiquitous and play an essential role in everyday life due to the tremendous diversity of their properties and to successful research aimed at manufacturing products with desired mechanical and chemical properties, tailored for each specific use. Therein, polymers have been investigated as components of composite materials, as parts of car bodies, sport equipment and many other household applications [1–3]. Recent advances have also been performed in biopolymers research to produce biodegradable and especially compostable packages, so they can act as fertilizers and soil conditioners [4].

The plasma modification of polymer surfaces has further increased their fields of applications. For instance, specific plasma treatments have been explored in the food packaging industry as a means to add missing barrier functionalities to polymer materials in addition to their intrinsic properties towards oxygen and water vapour permeation [5]. In medical surgery, plasma-treated polymer-composite materials such as poly(ethylene oxide terephthalate)–poly(butylene terephthalate) (PEOT/PBT), ultra-high-molecular-weight polyethylene (UHMWPE) and polytetrafluoroethylene(PTFE) have been widely used for applications such as bone fracture repair, joint replacements, dental bridges, vascular prosthesis with blood compatibility enhancement [6, 7].

Plasma treatments of fluoropolymer surfaces have drawn a special interest, particularly in the case of PTFE due to its outstanding properties: high thermal stability, lowest friction coefficient of all solids, hydrophobicity, chemical inertness in most types of environments, general inertness in the human body. In addition to its use for biocompatibility applications, PTFE has also been employed as self-cleaning coating [8–10]. Other widely used fluoropolymers such as PVDF and PFA have been plasma-treated to improve their chemical and mechanical properties [11–14].

Many articles are dedicated to the plasma treatment of fluoropolymer surfaces but most of them focus either on the surface characterization or on the plasma phase, without establishing the interfacial mechanisms that could link both. If the activation mechanisms on fluoropolymer surfaces are fairly well mastered, the understanding of the etching mechanisms occurring at the plasma/fluoropolymer interface is still not well established, particularly in atmospheric plasmas using $O_2$ as reactive gas [15, 16]. We propose here an experimental approach using the post-discharge of an RF atmospheric plasma torch supplied in helium and oxygen gases. Two etching models occurring at the post-discharge/polymer interface are then proposed after analyzing the results obtained from mass losses measurements, XPS spectra, WCA observations and AFM imaging.

## 2. Experimental set-up

### 2.1. Materials

In this work, we studied the high-density polyethylene (HDPE) polymer and five fluoropolymers: polyvinyl fluoride (PVF), polyvinylidene fluoride (PVDF), polytetrafluoroethylene (PTFE), fluorinated ethylene propylene (FEP) and perfluoroalkoxy (PFA), as represented in figure 1(a). All the polymer samples were purchased as foils from Goodfellow. After having been cut to 30×15mm2 in size, the samples were cleaned first in pure iso-octane and then in pure methanol, before being exposed to the plasma post-discharge.

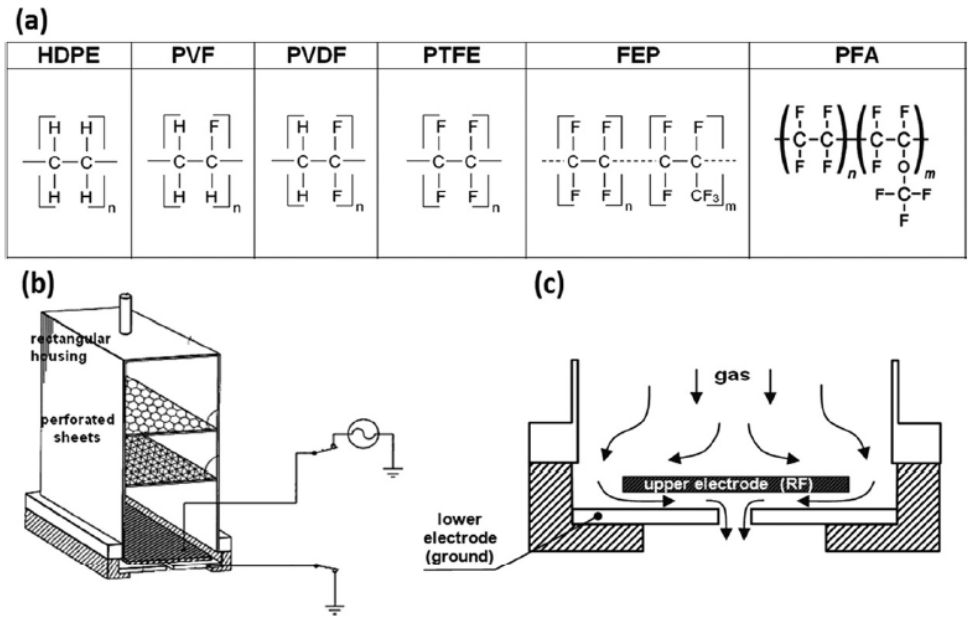

*Figure 1. (a) Presentation of the six polymers investigated in this work, (b), (c) cross sectional diagram of the RF-plasma source (Patent US7329608 Surfx Technologies [19]).*

### 2.2. Plasma torch

The sample surfaces were exposed to the post-discharge of an RF atmospheric plasma torch, the AtomfloTM 400L-Series, from SurfX Technologies [17]. The controller of this plasma source included an RF generator (27.12 MHz), an auto-tuning matching network and a gas delivery system with two mass flow controllers to regulate the helium and oxygen gases supplying the plasma source. Helium (vector gas) was studied for a flow rate fixed at 15 L.min$^{-1}$ and oxygen (reactive gas) for flow rates ranging from 0 to 200 mL.min$^{-1}$. The RF power commonly used was fixed at 90W. As presented in figure 1(b), the resulting gas mixture entered through a tube connected to a rectangular housing (55mm×20mm×80 mm). Inside, the gas was homogenized through two perforated sheets, then flew down around the left and right edges of the upper electrode and passed through a slit in the centre of the lower electrode. A plasma was struck and maintained between these electrodes by applying an RF power to the upper electrode while the lower electrode was grounded. The geometry of the slit was described as 'linear' due to the ratio of its aperture length (20 mm) to its width (0.8 mm). A background characterization of the mechanisms ruling this flowing post-discharge has already been published [18].

A robotic system was integrated to the plasma torch, enabling the treatment of large samples located downstream. The scanning treatment was achieved with respect to 3 degrees of freedom corresponding to the 3 axes of a Cartesian coordinate system. In all our experiments, the plasma source was only moved along one direction along a scan length of 30mm at a velocity of 25mm.s$^{-1}$.

## 2.3. Diagnostics

A drop shape analyser (Krüss DSA 100) was employed to measure dynamic contact angles of water drops deposited onto PTFE samples. As the static method based on the 'Sessile Drop Fitting' was shown to provide information that could not be easily interpreted (this approach was even called 'obsolete'), we measured advancing and receding contact angles by growing and shrinking the size of a single drop on the PTFE surface, from 0 to 20μl at a rate of 30μL.min$^{-1}$ [20]. The water contact angles (WCAs) plotted in this paper correspond to the advancing angles [21].

To evaluate the chemical composition at the surface of the samples, XPS analyses were performed with a Physical Electronics PHI-5600 instrument. The base pressure in the analytical chamber was ≈10$^{-9}$ mbar. Survey scans were used to determine the chemical elements present at the PTFE surface [22]. Narrow-region photoelectron spectra were used for the chemical study of the C 1s, O 1s and F 1s peaks. Spectra were acquired using the Mg anode (1253.6 eV) operating at 300W. Wide surveys were acquired at 93.9 eV pass-energy, with a 5 scans accumulation (time/step: 50 ms, eV/step: 0.8), and spectra of the C1s peaks at 23.5 eV pass-energy with an accumulation of 10 scans (time/step: 50 ms, eV/step: 0.025). The elemental composition was calculated after removal of a Shirley background line and using the sensitivity coefficients: $S_C$ = 0.205, $S_F$ = 1.000, $S_O$ = 0.63, $S_N$ = 0.38, $S_P$ = 0.25 and $S_{Al}$ = 0.11. The resulting compositions must be taken as indicative and are used only for comparison between the different plasma treatments (with/without O2). They do not reflect the absolute surface composition. In our case, several analysed regions on the same sample always led to the same relative compositions.

The surface morphology was further analysed with atomic force microscopy (AFM) which is well adapted to characterize the topographic profile of polymer surfaces at high lateral and vertical resolution [23]. In this work, all AFM images were recorded in air with a Nanoscope IIIa microscope operated in tapping mode (TM) [24]. The probes were commercially available silicon tips with a spring constant of 24–52N.m$^{-1}$, a resonance frequency lying in the 264–339 kHz range and a typical radius of curvature in the 5–10 nm range. The images presented in this article are height (5μm × 5μm) images recorded with a sampling resolution of 512×512 data points.

The samples were weighted before and after the plasma treatments to evaluate mass variations. For this, we used Sartorius BA110S Basic series analytical scales, characterized by a 110 g capacity and a 0.1 mg readability. Moreover, during the plasma treatment, every sample was placed on a large aluminum foil. As aluminum is known to be an efficient fluorine trap [25], we then analysed by XPS the hypothetical presence of polymer fragments on the foil.

Finally, the plasma phase was investigated by using two optical emission spectrometers: a first one for the visible range, a second one for the UV range. The study of the post-discharge in the visible range has already been published and all the details relative to the device as well [18]. To investigate the ultraviolet domain, we used a Shamrock 500i spectrometer from ANDOR, with a Czerny–Turner configuration, a focal length of 0.5 m, a reciprocal linear dispersion of about 1.5 nm.mm$^{-1}$ and a resolution close to 0.05 nm. We used a grating with 1800 lines.mm$^{-1}$, blazed at 250 nm. We chose an exposure time of 0.5 s and five accumulations.

## 3. Results

This section is dedicated to the influence of the number of scans (i.e. treatment time) performed by the plasma torch on the WCA measurements, the samples mass losses and the XPS spectra. Those techniques are also used with AFM to investigate the surface properties of polymers treated with a He–$O_2$ post-discharge by varying the $O_2$ flow rate. Finally, results dealing with the reactive species identified in the post-discharge will be recalled as they are involved in the etching mechanisms that we suggest in this article.

### 3.1. Influence of the number of scans (case of a He–$O_2$ post-discharge)

Polymer surface modifications were performed by setting the number of scans ($N_S$) of the plasma torch to different values (2, 10, 50, …, 2000). The torch-to-substrate distance (= gap) was fixed to 1 mm, the RF power to 90W, the helium flow rate to 15 L.min$^{-1}$ and the $O_2$ flow rate was 100 mL.min$^{-1}$. The advancing WCA are plotted versus $N_S$ in figure 2(a) for the six studied polymers: HDPE, PVF, PVDF, PTFE, FEP and PFA. For the sake of comparison, the advancing WCA measured on the native samples are indicated in figure 2(b). The correlation between these two graphs shows that the largest WCA variations are observed from the lowest $N_S$, i.e. the shortest treatment times. For instance, the dynamic WCA of HDPE decays from 95° to 22° in less than 10 scans and the same behaviour is noticed for PVF. Beyond this time interval, the trends remain almost constant. According to figure 2(a), two groups of polymers can be distinguished: those becoming more hydrophilic (HDPE, PVF and PVDF) and those becoming more hydrophobic (PTFE, FEP and PFA). The carbon backbone in the first group is entirely exposed to the reactive species of the post-discharge while a 'protective coating' of fluorine atoms surrounds the carbon skeletons of the polymers belonging to the second group. It seems that the higher the F/C ratio, the higher is the advancing WCA.

The receding WCA have not been plotted in figure 2(a) for the sake of clarity but also because they are known to be insensitive to the low-energy surface components [20, 21]. Receding WCA on HDPE or PVF substrates were not measurable due to the already low advancing WCA values. However, they were measured for the naturally hydrophobic surfaces such as PTFE, FEP and PFA. Figure 3(a) illustrates the hysteresis of dynamic WCA measured on a native PTFE surface. The inset pictures represent the advancing and receding WCA measured with the drop shape analyser at specific intervals expressed in seconds. This figure shows a hysteresis amplitude higher than 20° which was also representative of all the dynamic WCA measurements performed on the other hydrophobic polymers (PTFE, FEP and PFA). This elevated hysteresis amplitude attests the assumption of a Wenzel state where the water drop is in intimate contact with the solid asperities, as sketched in figure 3(b) [26–28].

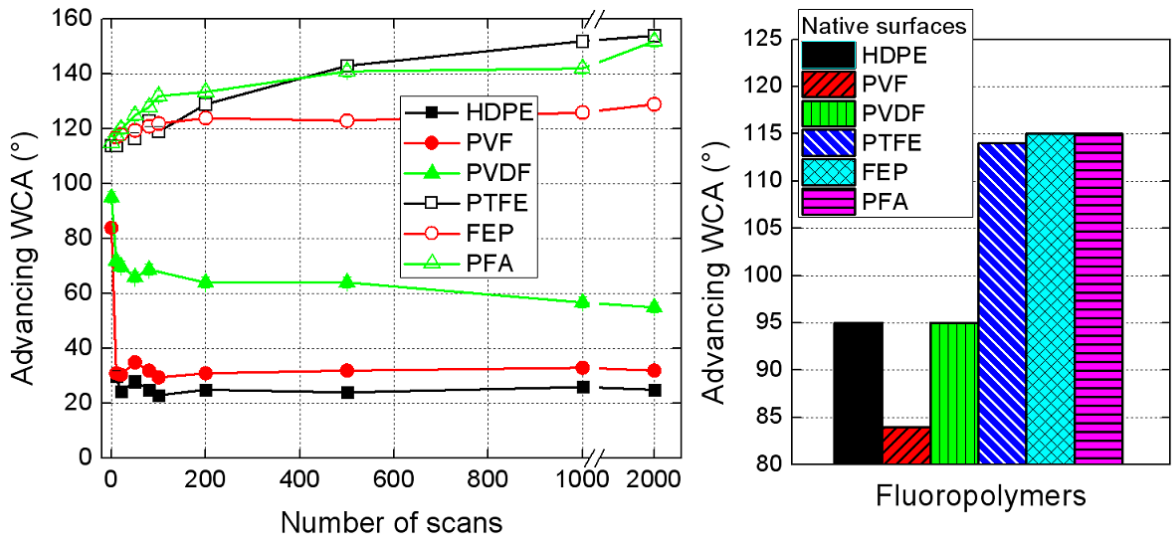

*Figure 2. (a) Advancing WCA measured versus the number of scans for a helium flow rate of 15 L.min$^{-1}$, an oxygen flow rate of 100 mL.min$^{-1}$, gap = 1mm and $P_{RF}$ = 90W, (b) advancing WCA for native polymer surfaces (HDPE, PVF, PVDF, PTFE, FEP and PFA).*

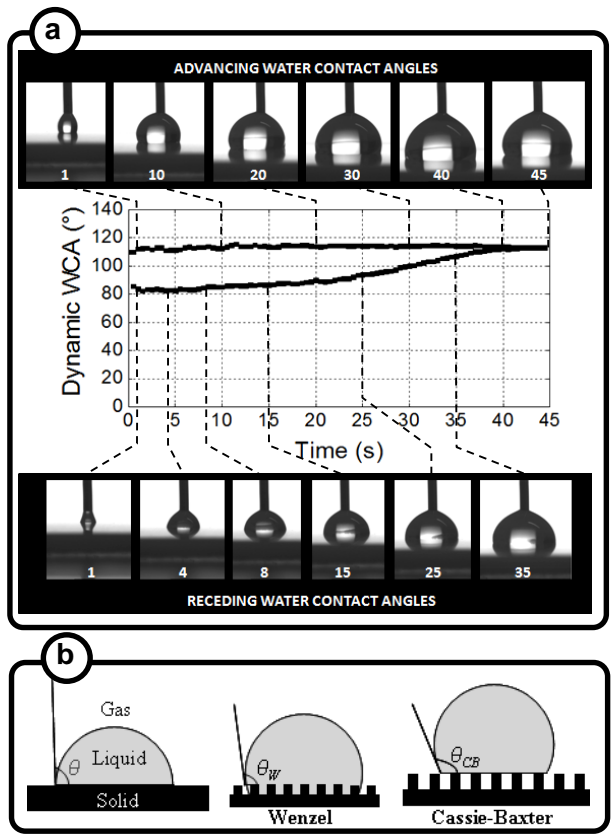

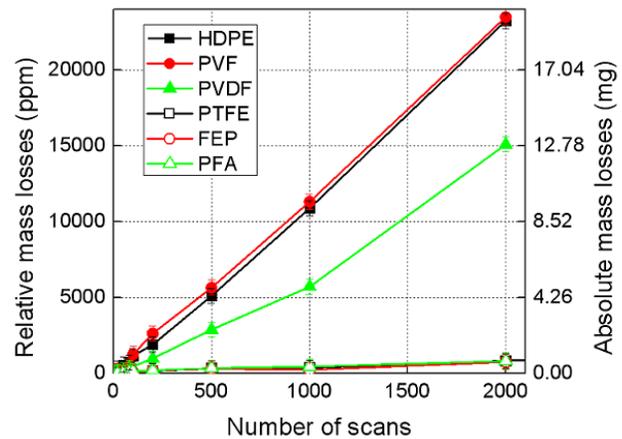

*Figure 4. Mass losses measured after plasma-treatment and expressed vs the number of scans for the same plasma treatment (helium flow rate = 15 L.min$^{-1}$, oxygen flow rate = 100 mL.min$^{-1}$, gap = 1mm and $P_{RF}$ = 90 W).*

*Figure 3. (a) Dynamic WCA obtained from a native PTFE surface. The inset pictures show the evolution of the drop deposition at specific intervals expressed in second, (b) representation of the Wenzel and Cassie-Baxter states.*

These WCA measurements were completed by mass measurements performed on the samples before and after plasma-treatment. The mass losses expressed in ppm and in mg are plotted in figure 4 versus the number of scans. On the whole range (from 0 to 2000 scans), the highest losses are found in the cases of HDPE and PVF, both following the same linear trend. For instance, for $N_S$ = 2000 scans, those two polymers lost almost 20 mg of their initial mass, while PVDF only lost 12.8 mg. For the three other fluoropolymers, namely PTFE, FEP and PFA, the mass loss is very small (about 0.6 mg). It thus appears that the higher the F/C ratio is, the lower the mass loss of the plasma-treated polymer is. This behavior is consistent with a former study by M A Golub et al which reported on the weight losses versus the exposure time for HDPE, PVF, PET, PVDF and Kapton exposed to ground state O ($^3$P) atoms downstream from a glow discharge [29].

The influence of $N_S$ (or the exposure time) can be interesting for industrial purposes if one desires to tailor surface properties with a specific hydrophobic character. From a fundamental point of view, studying the influence of the reactive gas is more challenging and might lead to an understanding of the mechanisms occurring between the polymer and the post-discharge at the macroscopic and the microscopic scales.

### 3.2. Influence of the $O_2$ flow rate

The influence of the $O_2$ flow rate on the advancing WCA is shown in figure 5(a) for the six polymer surfaces. The helium flow rate was maintained at 15 L.min$^{-1}$, $P_{RF}$ = 90W, $N_S$ = 1000 and the samples were placed 1mm away from the torch. The advancing angles for HDPE and PVF are similar and constant (around 22° ) for $O_2$ flow rates ranging between 0 and 200 mL.min$^{-1}$. Over the same range, PVDF gains in hydrophilicity but to a lower extent since the lowest advancing angle was 50°, as obtained for a $O_2$ flow rate as high as 200 mL.min$^{-1}$. A study by Vandencasteele et al showed that the treatment of PVF and PVDF surfaces by a $O_2$ discharge at low pressure, led to similar WCA for the treated PVF surfaces (almost 25°) whereas lower WCA were obtained for the PVDF surfaces (around 10°) [39]. In this last case, the reason can be attributed to the highest content of oxygenated functions on the PVDF surface than in the present case (22% for Vandencasteele et al, while 11.2% in our study).

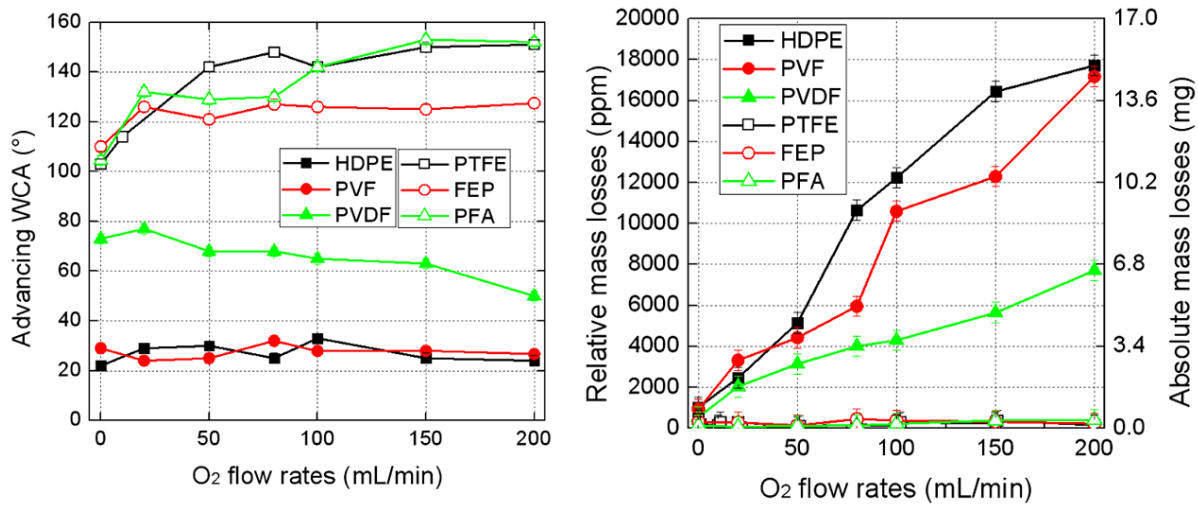

Figure 5. (a) Advancing WCA of HDPE, PVF, PVDF, PTFE, FEP and PFA measured versus the $O_2$ flow rate for a helium flow rate set at 15 L.min$^{-1}$, $N_S$ = 1000, gap = 1mm and $P_{RF}$ = 90W, (b) Mass loss measurements of the six polymer surfaces versus the $O_2$ flow rate for a helium flow rate set at 15 L.min$^{-1}$, $N_S$ = 1000, gap = 1mm and $P_{RF}$ = 90W.

The trends for the three other polymers (PTFE, FEP and PFA) are the same: an increase in the advancing WCA versus the reactive gas flow rate is observed, with a maximum value of 150° reached at 200 mL.min$^{-1}$ of $O_2$. In the literature, the treatment of PFA has already been performed by helium or oxygenated low pressure discharges resulting in WCA decaying around 80° [30]. Such a decrease in the hydrophobicity was due to a defluorination and oxidation of the PFA surface which is not observed here, probably thanks to the milder conditions offered by an atmospheric post-discharge. The treatment of FEP surfaces by an argon or an oxygen low pressure discharge led also to a decrease in the hydrophobicity, with WCA of 105° and 90°, respectively [31]. Another team also treated FEP and PFA films but at atmospheric pressure with a plasma glow supplied either in He-$O_2$ or in pure helium [11]. For both conditions, the WCA of the polymers decreased to around 65°. The plasma treatments of PTFE at atmospheric pressure most often lead to a decrease of WCA [32–34]. Therefore, the increase in WCA that is measured in this work appears more related to the post-discharge conditions than to the atmospheric pressure, and particularly to the scanning mode, which allows surface etching without heating the whole sample too much (not higher than 400 K).

Mass losses measurements also evidenced the influence of the $O_2$ flow rate, as illustrated in figure 5(b). Polymers of the first group show the same linearly increasing mass losses with values reaching 15.2 mg for the largest $O_2$ flow rate. In contrast, polymers belonging to the second group undergo vanishingly small mass losses, slightly decreasing from less than 1mg to 0 mg. These two radically different behaviors are detailed in the discussion.

Measurements of WCA and mass losses were completed by a chemical surface analysis by XPS. The results are summarized in table 1. In figures 6 and 7, we show wide survey spectra and narrow-region spectra for HDPE and PVDF; some of the other peaks have been published previously [10, 35]. In table 1, the grey columns show the relative atomic compositions and the white columns the relative distribution in the functions detected on the various polymer surfaces (HDPE, PVF, PVDF, PTFE, FEP and PFA) but also on the aluminum foils placed closed to these surfaces during the plasma treatments. Therefore, twelve cases have to be considered and for each one, table 1 reports the relative surface composition of the native surface (row 'native'), the surface treated by a helium post-discharge (row 'He') and the surface treated by a helium–oxygen post-discharge (row 'He-$O_2$'). The binding energies of the identified functions can easily be found in the literature [30, 31, 36–39].

| | | C | O | F | N | P | Al | C-C or C-H | C-OH | O-C-O or C=O | O-C=O | CHF-CH2 | CH2-CF2 | CH2-CHF-CH2 | CF2-CHF or CF2-CF | CF | CF2 | CF3 |
|---|---|---|---|---|---|---|---|---|---|---|---|---|---|---|---|---|---|---|
| HDPE | native | 100 | 0 | 0 | 0 | 0 | 0 | 100 | 0 | 0 | 0 | 0 | 0 | 0 | 0 | 0 | 0 | 0 |
| | He | 66.7 | 29 | 0 | 1.9 | 2.4 | 0 | 64.6 | 12.9 | 6.5 | 16 | 0 | 0 | 0 | 0 | 0 | 0 | 0 |
| | HeO₂ | 58.4 | 34.6 | 0 | 1.5 | 5.5 | 0 | 57.7 | 13.1 | 12.3 | 16.9 | 0 | 0 | 0 | 0 | 0 | 0 | 0 |
| PVF | native | 75.2 | 3.6 | 21.2 | 0 | 0 | 0 | 61.9 | 0.5 | 0 | 3 | 34.6 | 0 | 0 | 0 | 0 | 0 | 0 |
| | He | 65.3 | 19.4 | 12.3 | 1.5 | 1.5 | 0 | 51.5 | 1.2 | 1.2 | 12.2 | 33.9 | 0 | 0 | 0 | 0 | 0 | 0 |
| | HeO₂ | 65.8 | 20.9 | 13.3 | 0 | 0 | 0 | 37.3 | 10 | 8.2 | 13.4 | 31.1 | 0 | 0 | 0 | 0 | 0 | 0 |
| PVDF | native | 61.9 | 2.3 | 35.2 | 0 | 0 | 0 | 53.8 | 0.8 | 0.7 | 0 | 0 | 44.7 | 0 | 0 | 0 | 0 | 0 |
| | He | 58.5 | 11.4 | 30.1 | 0 | 0 | 0 | 37.1 | 6.9 | 3 | 1.1 | 3.5 | 35.8 | 0 | 0 | 0 | 9.1 | 3.5 |
| | HeO₂ | 57.2 | 11.2 | 31.6 | 0 | 0 | 0 | 39.7 | 5.2 | 3.5 | 1.1 | 1.6 | 43.2 | 0 | 0 | 0 | 4.1 | 1.6 |
| PTFE | native | 41.4 | 0.3 | 58.3 | 0 | 0 | 0 | 11.3 | 0 | 0 | 0 | 0 | 0 | 0 | 0 | 0 | 88.7 | 0 |
| | He | 39.9 | 1.8 | 58.2 | 0 | 0 | 0 | 8.3 | 0 | 0 | 0 | 0 | 6.6 | 6.8 | 0 | 0 | 78.3 | 0 |
| | HeO₂ | 43.7 | 1.0 | 55.3 | 0 | 0 | 0 | 7.4 | 0 | 0 | 0 | 0 | 4.7 | 3.2 | 0 | 0 | 84.7 | 0 |
| FEP | native | 42.9 | 0 | 56.4 | 0.7 | 0 | 0 | 9.7 | 4.8 | 0 | 0 | 0 | 0 | 0 | 0 | 5.9 | 74.1 | 5.5 |
| | He | 37.1 | 0 | 62.9 | 0 | 0 | 0 | 10.1 | 4.1 | 0 | 0 | 0 | 0 | 0 | 0 | 5.1 | 76.1 | 4.6 |
| | HeO₂ | 37.6 | 0.5 | 61.9 | 0 | 0 | 0 | 7.8 | 3.4 | 0 | 0 | 0 | 0 | 0 | 0 | 5.2 | 77.9 | 5.7 |
| PFA | native | 41.7 | 0.7 | 57.6 | 0 | 0 | 0 | 15.7 | 2 | 0 | 0 | 0 | 0 | 0 | 0 | 3.9 | 74.7 | 3.7 |
| | He | 44.2 | 1.4 | 54.5 | 0 | 0 | 0 | 6.3 | 4 | 5.2 | 0 | 0 | 0 | 0 | 0 | 8.5 | 72.1 | 3.9 |
| | HeO₂ | 41.7 | 2 | 56.3 | 0 | 0 | 0 | 7.2 | 2.3 | 1.3 | 0 | 0 | 0 | 0 | 0 | 2.5 | 80.9 | 5.8 |
| Al foil HDPE | native | 47.3 | 29.5 | 0 | 0 | 0 | 23.2 | 85.7 | 3.1 | 3.2 | 8 | 0 | 0 | 0 | 0 | 0 | 0 | 0 |
| | He | 38.6 | 35.9 | 0 | 4 | 0 | 21.5 | 56.8 | 11.7 | 7.5 | 24 | 0 | 0 | 0 | 0 | 0 | 0 | 0 |
| | HeO₂ | 39.7 | 42.5 | 0 | 3.4 | 0 | 14.4 | 41.9 | 12 | 6.2 | 39.9 | 0 | 0 | 0 | 0 | 0 | 0 | 0 |
| Al foil PVF | native | 47.3 | 29.5 | 0 | 0 | 0 | 23.2 | 85.7 | 3.1 | 3.2 | 8 | 0 | 0 | 0 | 0 | 0 | 0 | 0 |
| | He | 40.2 | 32.8 | 3.4 | 5.8 | 0 | 17.8 | 55.0 | 11.8 | 7.8 | 25.4 | 0 | 0 | 0 | 0 | 0 | 0 | 0 |
| | HeO₂ | 29.3 | 37.8 | 8.7 | 3.1 | 0 | 21.2 | 44.6 | 13.1 | 10.2 | 32.1 | 0 | 0 | 0 | 0 | 0 | 0 | 0 |
| Al foil PVDF | native | 47.3 | 29.5 | 0 | 0 | 0 | 23.2 | 85.7 | 3.1 | 3.2 | 8 | 0 | 0 | 0 | 0 | 0 | 0 | 0 |
| | He | 27.3 | 22.8 | 18.6 | 5.5 | 0 | 25.8 | 57.5 | 17.1 | 6.4 | 19 | 0 | 0 | 0 | 0 | 0 | 0 | 0 |
| | HeO₂ | 22.5 | 22.1 | 24.3 | 3.1 | 0 | 27.9 | 50.2 | 17.7 | 7.2 | 24.9 | 0 | 0 | 0 | 0 | 0 | 0 | 0 |
| Al foil PTFE | native | 47.3 | 29.5 | 0 | 0 | 0 | 23.2 | 85.7 | 3.1 | 3.2 | 8 | 0 | 0 | 0 | 0 | 0 | 0 | 0 |
| | He | 32.8 | 7.6 | 48.4 | 0 | 0 | 11.2 | 6.6 | 5.4 | 4.3 | 6.9 | 0 | 0 | 0 | 0 | 0 | 76.8 | 0 |
| | HeO₂ | 27.8 | 30.8 | 11.9 | 0 | 0 | 29.6 | 57.4 | 7.4 | 3.3 | 10.6 | 0 | 0 | 0 | 0 | 0 | 21.3 | 0 |
| Al foil FEP | native | 47.3 | 29.5 | 0 | 0 | 0 | 23.2 | 85.7 | 3.1 | 3.2 | 8 | 0 | 0 | 0 | 0 | 0 | 0 | 0 |
| | He | 43 | 2.6 | 53 | 0 | 0 | 1.4 | 8.1 | 8.1 | 5.3 | 7.4 | 0 | 0 | 0 | 0 | 0 | 63.9 | 7.2 |
| | HeO₂ | 22.8 | 24 | 25.8 | 0 | 0 | 27.4 | 53.6 | 11.3 | 1.7 | 7.3 | 0 | 0 | 0 | 0 | 0 | 26.1 | 0 |
| Al foil PFA | native | 47.3 | 29.5 | 0 | 0 | 0 | 23.2 | 85.7 | 3.1 | 3.2 | 8 | 0 | 0 | 0 | 0 | 0 | 0 | 0 |
| | He | 26.2 | 14.4 | 38.1 | 2.2 | 0 | 19.1 | 16.8 | 8.9 | 5.7 | 11.2 | 0 | 0 | 0 | 0 | 0 | 55.4 | 1.8 |
| | HeO₂ | 21.2 | 24.8 | 30 | 1.1 | 0 | 22.8 | 45.4 | 9 | 2.7 | 7.5 | 0 | 0 | 0 | 0 | 0 | 35.4 | 0 |

*Table 1. Relative surface composition of polymer surfaces and aluminum foils placed closed to them, for three cases (native surface, surface treated by a He post-discharge and treated by a He–O₂ post-discharge).*

The survey of the native HDPE polymer is represented by the black curve in figure 6. It clearly shows that the surface is only composed of carbon (and hydrogen) (100%). No oxygenated functions or other contaminants are observed, which reflects the efficiency of the cleaning process. The treatment of the same surface by a He post-discharge (and even more by a He-O$_2$ post-discharge) induced the emergence of a significant O 1s peak pointing out the incorporation of oxygenated functions, as illustrated by the narrow-region C 1s spectra in figures 7(b) and (c). The oxygenated functions detected in the case of the He post-discharge are attributed to the dissociation of the ambient O$_2$ by interacting with the post-discharge. In both cases, the decomposition of the C 1s peak demonstrates the existence of four components: C–C (or C–H), C–OH, O–C–O (or C=O) and O–C=O, appearing at the following binding energies 285.0 eV, 286.5 eV, 288.0 eV and 289.5 eV [40]. These four components were also detected by decomposing the C 1s peaks measured on the aluminum foils (see table 1). For instance, the relative composition of the aluminum foil in C–OH components increased from 3.1% (native) to 12% (He-O$_2$ post-discharge). As reported in table 1, the surface of the native aluminum foil was aluminum oxide, which is spontaneously formed at ambient air.

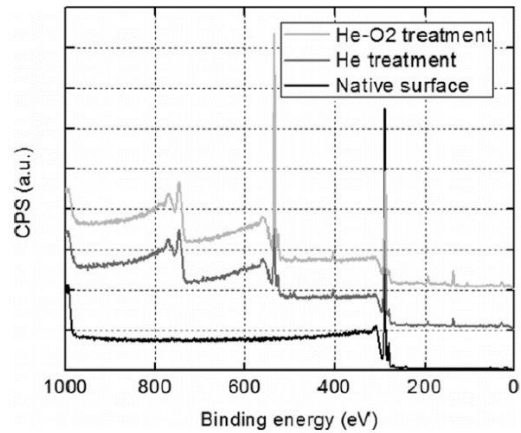

Figure 6. XPS survey spectra obtained for a native surface of HDPE and for plasma-treated HDPE surfaces with a helium or a Helium-oxygen post-discharge (He flow rate = 15 L.min$^{-1}$ and O$_2$ flow rate = 200 mL.min$^{-1}$).

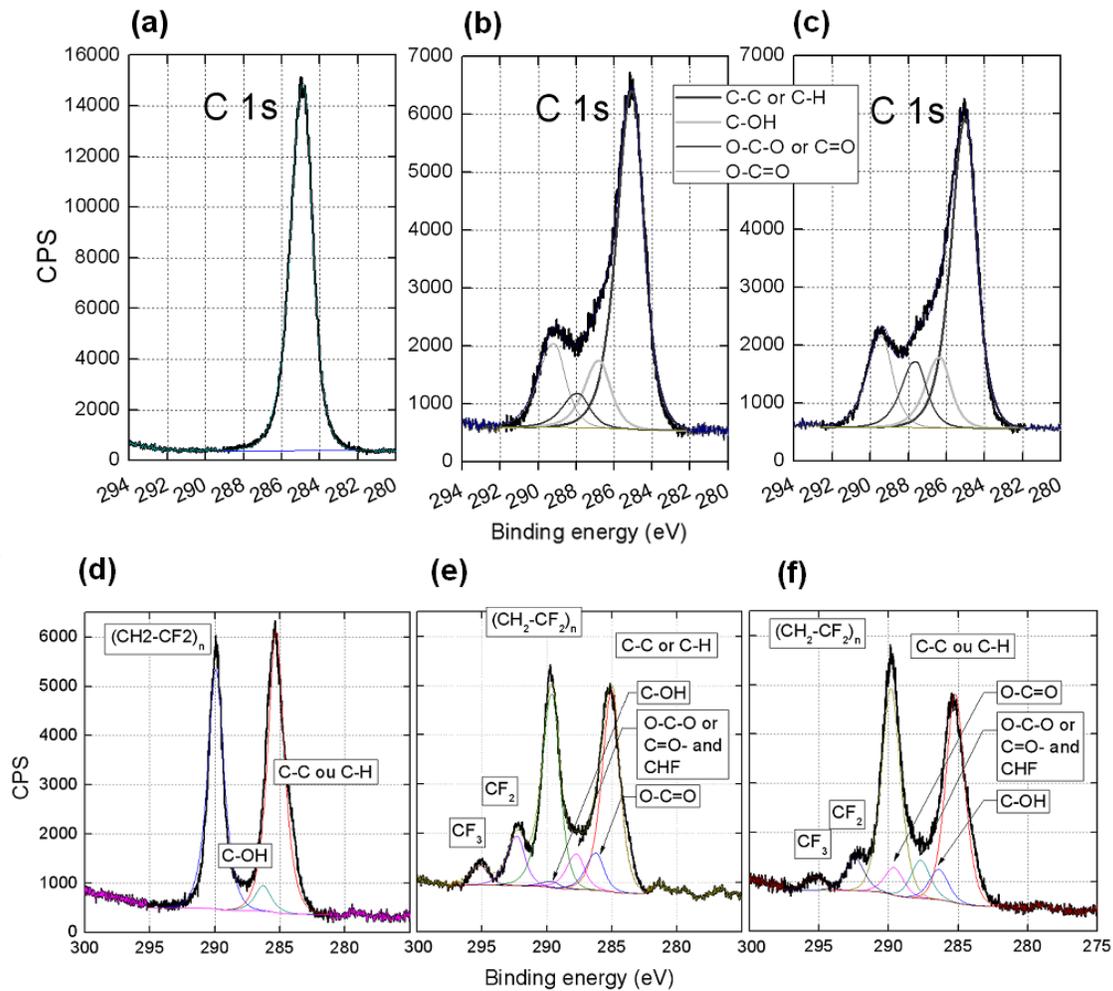

Figure 7. Narrow-region C 1s spectra obtained for HDPE and PVDF surfaces (a) native HDPE, (b) HDPE treated by a He post-discharge, (c) HDPE treated by a He-O$_2$ post-discharge, (d) native PVDF, (e) PVDF treated by a He post-discharge and (f) PVDF treated by a He-O$_2$ post-discharge.

Although very weak, some contamination peaks (nitrogen and phosphorus) were also detected and may come either from the quality of the gases used for these experiments or from heat sensitive materials located inside the plasma torch and likely to molder after a long time of operation. These contaminants only appeared for the treatment of HDPE surfaces and are anyway irrelevant for the discussion.

As expected, the native PVF and PVDF samples present a lower carbon content (75.2% and 61.9%, respectively). Figure 7(d) represents the narrow region C 1s spectrum of a native PVDF surface. It indicates the two main components C–C (or C–H) and $CH_2$–$CF_2$ centered at 285.0 eV and 291.0 eV. Moreover, the presence of a slight oxidation was detected at 286.5 eV corresponding to the C–OH component. According to figures 7(e) and (f), the helium and helium–oxygen plasma treatments both lead to the same oxygenated functions, already detected in the case of a treated HDPE surface. Furthermore, fluorinated components, namely CHF–$CH_2$ and $CF_2$ were evidenced at 288.0 eV and 292.6 eV, respectively, and also a component close to 295 eV (approximately 294.6 eV), which was assigned to the $CF_3$ function as $CF_3$ appears at 294.6 eV in the case of an FEP polymer [31].

In the case of the PTFE, FEP and PFA samples, the carbon content is around 42%. An elaborate study reporting the effect of He and He-$O_2$ post-discharges treatments has been led by Hubert et al [35]. Globally, the plasma treatment of these surfaces, whatever the O2 flow rate, allows the formation of oxygenated functions for the first group of polymers and fluorinated functions (such as CHF–$CH_2$, $CF_2$) for the polymers belonging to the second group.

Finally, we performed AFM pictures on two polymer surfaces: HDPE (belonging to the group of polymers becoming more hydrophilic, including therefore PVF and PVDF) and PTFE (belonging to the group of polymers becoming more hydrophobic, including also FEP and PFA). For each polymer surface, figure 8 mentions the root-mean square (RMS) roughness values corresponding to the native surface and to the two plasma treatments previously described. Whatever the polymer, the helium treatment induces a decrease in the roughness value. This effect is more pronounced in the case of HDPE, for which the native surface exhibits an average roughness of 51.1 nm while the helium-treated surface has a roughness of 19.6 nm. The decrease is less significant for the PTFE samples (32.0 nm versus 27.7 nm). In contrast, for the helium–oxygen treatment, we observe an increase of the roughness values compared to the native surfaces. Here also, the effect is more pronounced for HDPE samples: the roughness increases from 51.1 nm up to 79.1 nm for HDPE while it only increases from 32.0 to 38.2 nm for PTFE.

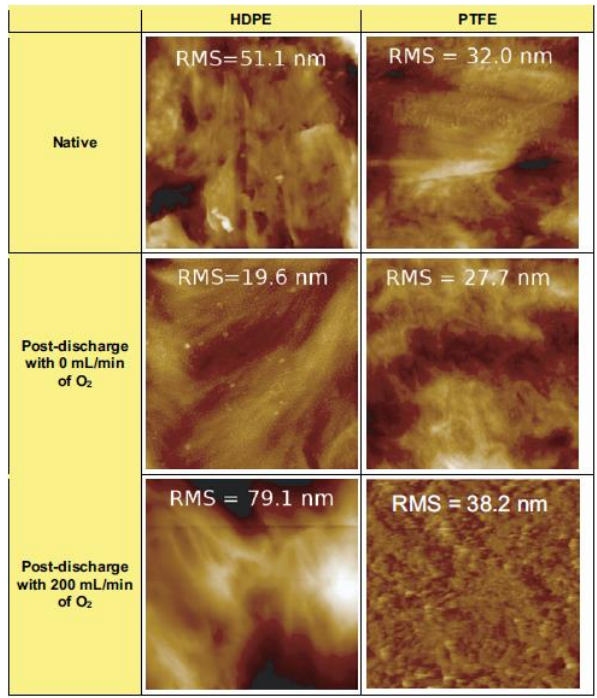

*Figure 8. Tapping-mode AFM images (5µm × 5µm) of HDPE and PTFE surfaces before and after plasma treatments performed by the RF-plasma torch at atmospheric pressure.*

As we focus on the mechanisms occurring at the polymer/plasma interface, our surface analyses were completed by a characterization of the post-discharge. This study was performed by mass spectrometry and optical emission spectroscopy (in the visible range). Our recent results have highlighted the presence of helium metastable species in a pure helium post-discharge. They also showed that the addition of a low $O_2$ flow rate to the carrier gas was consuming those metastable He $2^3S$ species to produce $O_2^+$ and $N_2^+$ ions from Penning ionizations and subsequently O radicals and $O_2$ metastable molecules [18]. No emission of $O^+$ ions was evidenced in the post-discharge:

$$\text{He}(2^3S) + O_2 \rightarrow \text{He} + O_2^+ + e,$$

$$\text{He}(2^3S) + N_2(X\ ^1\Sigma_g^+, v=0) \rightarrow \text{He} + N_2^+(B\ ^2\Sigma_u^+, v'=0) + e$$

In this paper we also evidenced by optical absorption spectroscopy the presence of the He metastable species in a pure helium post-discharge. We also suggested the fact that their lifetime was sufficiently elevated for allowing them to reach the surface of the sample. We also showed that the $O_2$ metastable species were detected in the He–O2 post-discharge by tracking the optical emission of $O_2$ ($^1\Sigma_g^+$) species. Our experiments were consistent with the lifetimes of the He $2^3S$ and the $O_2^M$ species respectively evaluated by Tachibana et al and Jeong et al under similar conditions [41, 42].

The collisions and radiative processes of a helium post-discharge operating at atmospheric pressure are dominated by step-wise processes (i.e. the excitation of an already excited atomic/molecular state) and by 3-body collisions leading to the formation of excimers [43–45]. The assumption of helium excimers produced in the discharge (i.e. between the electrodes) is reasonable, based on the presence of He metastable species evidenced by optical absorption spectroscopy. By investigating the VUV domain of the pure He discharge with the Andor spectrometer, we found a He$_2$* excimer fluorescence at 60.2 nm, as represented in figure 9(a). This fluorescence of He$_2$∗ at 60.2 nm is named 'first continuum'. A 'second continuum' is also present between 60 and 100 nm but it could not be observed in our experimental conditions due to its low emissivity [44]. Figure 9(b) represents the evolution of the first continuum intensity versus the $O_2$ flow rate in the post-discharge. The highest excimer emission is obtained for a pure He post-discharge and it decays when $O_2$ is mixed with the carrier gas. Therefore, the evolution of the excimer emission follows the same trend as the He metastable emission [18]. Moreover, the excimer lifetime is supposed to not overpass 10µs since this value corresponds to the lifetime of the He metastable species. In other words, the excimer species are dissociated before reaching the polymer surface, while the excimer VUV radiation can interact with the surface. As the post-discharge can be considered as an optically very thin medium, this VUV radiation is weakly absorbed.

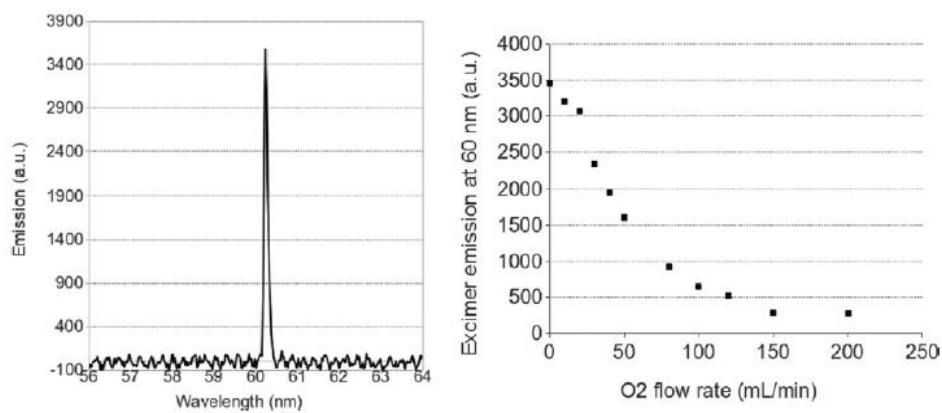

*Figure 9. (a) Optical emission spectrum of the first $He_2^*$ excimer continuum measured in a pure He post-discharge at atmospheric pressure (15 L.min$^{-1}$, 120 W). (b) Variation of the excimer emission at 60 nm versus the $O_2$ gas mixed with the carrier gas.*

# 4. Discussion

## 4.1. Representations of the etching mechanisms occurring at the post-discharge / fluoropolymer interface

The correlation of all the previous results led us to propose two etching models represented in figure 10. Model A applies to a polymer whose carbon backbones are entirely exposed to the post-discharge; it is suitable for polyethylene (HDPE, LDPE). Model B is relevant for fluoropolymers whose carbon backbones are entirely protected from the post-discharge by the fluorine atoms. It applies to PTFE, FEP and PFA polymers. As the C–C chains from the PVF and PVDF polymers are partially covered by fluorine atoms, they can be regarded as intermediate cases between models A and B. For each model, two cases are treated: a post-discharge only supplied with helium gas (15 L.min$^{-1}$) and a post-discharge supplied with a helium-oxygen gas mixture (15 L.min$^{-1}$ for He and 200 mL.min$^{-1}$ for $O_2$). Four cases can thus be distinguished and are detailed in each of the following sections.

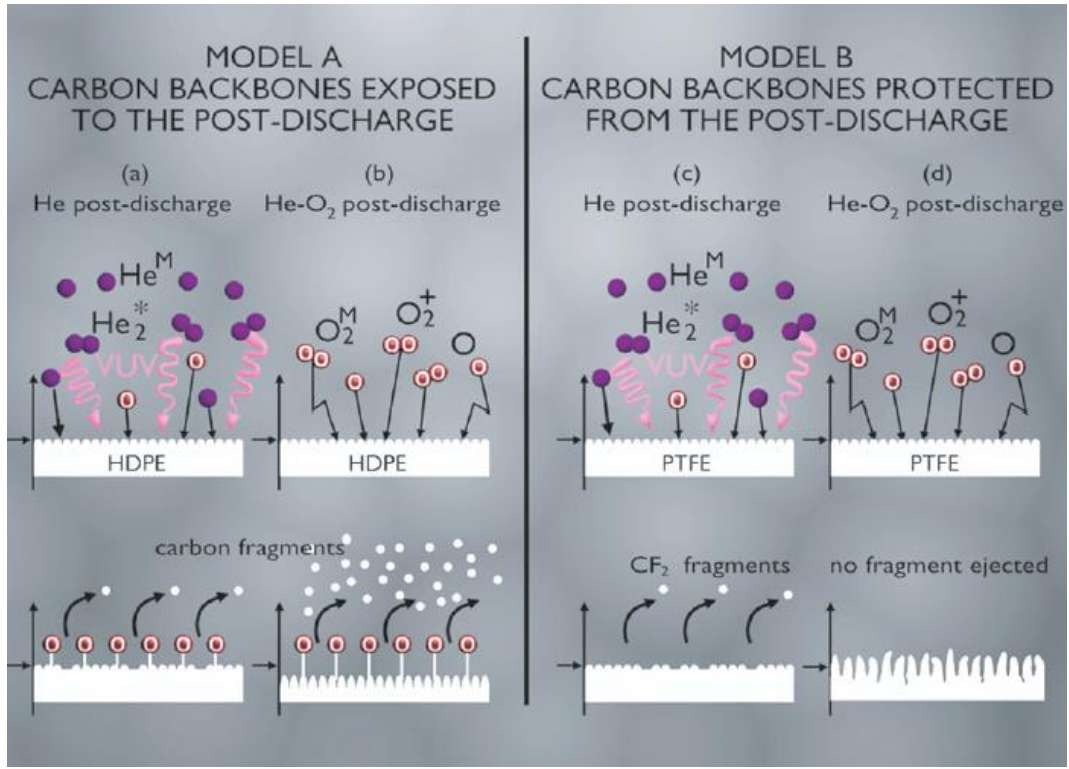

Figure 10. Models representing the etching mechanisms occurring on the HDPE and PTFE surfaces, either by a pure helium post-discharge or by a helium–oxygen post-discharge at atmospheric pressure.

## 4.2. HDPE surface/He post-discharge

**4.2.1. Macroscopic scale**

The XPS spectra evidenced the formation of C–OH, O–C–O and O–C=O groups on the HDPE surface treated by the helium post-discharge (see table 1). These oxygenated functions can only be incorporated on the surface by breaking some C–C and/or C–H bonds, and explain the decrease in the hydrophobicity evidenced by the low values of dynamic WCA. XPS measurements performed on the aluminum foil attest low changes of its surface chemical composition, thus discarding the assumption of numerous polymer fragments being ejected during the treatment. Besides, this result is consistent with the sample weightings performed before and after the treatment, showing very mass losses (around 0.5 mg). As the AFM images indicate a significant decrease in the roughness while the XPS spectra showed an increase in the oxygenated functions, a synergistic effect between the surface smoothening and its oxidation is achieved to decrease the hydrophobicity.

**4.2.2. Microscopic scale**

The bond dissociation energies (BDE) of C–C, C–H, C–F, O–H and C–O are reported in tables 2 and 3. The BDE of the bond between two C atoms in $sp^3$ hybridization is 3.55 eV. This is so in such widely different molecules as ethane, polyethylene and diamond [46, 47]. Table 3 indicates that the C–H bond can be dissociated at slightly different energies depending on the environment of the carbon atom (i.e.

primary, secondary or tertiary carbon), on resonance effects, on steric strain induced by bulky neighbouring groups and on the rigidity of their own or adjacent valence structures [48, 49]. Globally, it appears that the lower the H/C ratio, the lower C–H BDE. In the case of polyethylene, the BDE of C–H is 3.8 eV. If it is easier to break a C–C bond than a C–H bond, it does not mean that more C atoms are ejected from the surface than H atoms. Indeed, the ejection of one H atom requires only the breaking of a C–H bond while the ejection of one C atom requires at least the simultaneous breaking of two C–C bonds. For this reason, the amount of C atoms ejected is quite low while the amount of ejected H atoms is much more elevated, thus allowing the formation of C–O bonds on the HDPE surface. Some other works are related to the exposure of HDPE surfaces to an inert gas plasma (He, Ar, etc.) [50–52]. In those cases, the treatments were sufficient to abstract H atoms and to form free radicals at or near the surface. Those radicals could then interact to form the cross-links and unsaturated groups with the chain scission [53].

| Bond | BDE (kJ mol$^{-1}$) | BDE (eV) | Ref. |
|---|---|---|---|
| C–F | 452–531 | 4.5–5.3 | [58] |
| O–H | 454 | 4.5 | [59] |
| C–O | 400 | 4.0 | [59] |
| C–H | 368 | 3.7 | [48] |
| C–C | 355 | 3.6 | [47] |
| O=O | 528 | 5.3 | [59] |

Table 2. Bond dissociation energies. The conversions between kJ.mol$^{-1}$ and eV are performed with the approximation 1 kJ.mol$^{-1}$ = 1.038.10$^{-2}$ eV.

| | H/C Ratio | BDE (kJ mol$^{-1}$) | BDE (eV) | Carbon type | Ref. |
|---|---|---|---|---|---|
| CH$_3$CH$_3$ | 3 | 422 | 4.2 | Primary | [60] |
| CH$_3$CH$_2$-H | 3 | 420 | 4.2 | Primary | [61] |
| (CH$_3$)$_2$CH-H | 2.7 | 409 | 4.1 | Secondary | [61] |
| CH$_3$CH$_2$CH$_3$ | 2.6 | 411 | 4.1 | Secondary | [60] |
| (CH$_3$)$_3$CH | 2.5 | 404 | 4.0 | Tertiary | [60] |
| Polyethylene | 1 | 384 | 3.8 | Secondary | [47] |

Table 3. C–H bond dissociation energies for various aliphatic hydrocarbons at 298 K.

Due to their too short lifetime, the He$_2$* excimers cannot reach the polyethylene surface. Even if the excimers dissociate to generate excited helium species (He*) carrying enough energy to break C–H and C–C bonds, their noble gas electron configuration (1s$^2$) prevents them from reacting with a C* radical issued from a bond breaking [54]. The only active species assumed to react with the C* are: (i) the O radicals coming from the O$_2$ of the air and subsequently dissociated by the post-discharge. After breaking a C–H (or a C–C) bond, the O radicals can substitute the H atom since the C–O BDE (4.0 eV) is slightly higher than that of C–H (3.8 eV). (ii) The He metastable species. Due to their long lifetime they can reach the surface of the sample to transfer their potential energy. However, no 'sputtering' of the surface is expected since the atomic mass of a helium metastable species is too low. No VUV radiation of the helium metastable species was detected probably due to their involvement in the excimer production processes which operate on shorter timescales than the radiative processes (iii) The VUV photons emitted by He$_2$* excimers from the first continuum at 60 nm (20.7 eV). Hydrocarbon polymers are characterized by a very strong absorption below 160 nm, which originates from the dissociative excitation of the C–C and C–H bonds [55, 56]. For this reason, the VUV radiation is assumed to be responsible for the surface smoothening evidenced by AFM in the absence of O$_2$ in the discharge. This assumption has already been raised by Fricke et al in the case of an Ar/O$_2$ atmospheric plasma jet [57].

## 4.3. HDPE surface/He–O$_2$ post-discharge

**4.3.1. Macroscopic scale**

As in the previous case, the surface is activated by oxygenated functions and their relative abundance is slightly higher due to the addition of O$_2$ in the post-discharge (see figure 10(b)). For this reason, a high hydrophilicity state is reached, with dynamic WCA decreasing to 22°. As in a pure helium post-discharge, the production of these functions can only be achieved by breaking some C–C and/or C–H covalent bonds. Here, the mass losses are much more elevated (reaching almost 17 mg) while they can be considered as negligible (lower than 1 mg) when no $O_2$ was added to the post-discharge. Consequently, the chemical composition of the aluminum foil surface is changed, indicating that the ejected fragments present a higher content in oxygenated functions.

The AFM data indicate an increase in the roughness, from a value of 51.1 nm for the native surface to 79.1 nm after the He-$O_2$ post-discharge treatment. An increase in the roughness is usually correlated with an increase in the hydrophobicity provided that no functionalization of the surface is operated. Contrary to the 'HDPE/helium post-discharge' treatment, the roughening effect is counterbalanced by the oxidation effect. Instead of being in synergy, they are here in competition and since the dynamic WCA remain as low as before (about 22°), the oxidation effect is assumed to be stronger than the roughening effect. Similar results have already been obtained by Lehocky et al in the case of an air and an oxygen low pressure RF-plasma treatment. Polyethylene surfaces with a stronger hydrophilic character were obtained, mainly due to the generation of carboxylic, carbonyl and peroxide groups at the surface [62]. Our results are also consistent with those from Guruvenket et al on HDPE exposed to an ECR glow discharge: oxygen functional groups were evidenced at the surface, including C–O, C=O, O–C=O [53]. Here again, a balance of two processes was highlighted: (i) the formation of oxygen functional groups at the polymer surface and (ii) the polymer surface etching through the reaction of atomic oxygen with the surface carbon atom, resulting in volatile reaction products. An important increase in the roughness due to the plasma etching and restructuring was also evidenced [62].

**4.3.2. Microscopic scale**

In the post-discharge, most of the He metastable species have been consumed by Penning ionizations, thus preventing the production of helium excimers and therefore of VUV radiation. The major active species interacting with the polyethylene surface are the O and OH radicals but also the $O_2$ metastable molecules and the $O_2^+$ ions. As the emission of the O and OH radicals is higher than in a pure He post-discharge, the densities of these species are also considered as being higher in the present case [18]. They are responsible for the massive ejection of polymer fragments (15 mg here while it is lower than 1mg in the case of a He post-discharge).

The $O_2^M$ and $O_2^+$ species are also assumed to participate to the etching even if their densities are very small. Their momentum is indeed twice higher than that of an O radical, thus enabling them to break C–C and C–H covalent bonds more efficiently. The dissociation of a C–C bond (3.7 eV) by a $O_2^M$ ($b^1\Sigma_g^+$) species is expected to induce the dissociation of the $O_2^M$ into O radicals. Indeed, the $O_2^M$ BDE is comprised between 1.7 and 2.8 eV depending on the inter-nuclear distance, for the firsts (v = 0–5) vibrational levels [63]. The important role played by the oxygen radicals on the etching mechanisms has already been reported by Fricke et al in the case of polymethyl methacrylate (PMMA), PE, PP, polycarbonate (PC) and polystyrene(PS) [57]. Due to the high reactivity of atomic oxygen, the two following reactions occur:

$$RH + O \rightarrow R\cdot + \cdot OH \qquad (3)$$
$$RH + \cdot OH \rightarrow R\cdot + H_2O \qquad (4)$$

The authors also explain that the surface oxidation can lead to the formation of alkoxy and peroxy radicals:

$$R\cdot + O \rightarrow RO\cdot \qquad (5)$$
$$R\cdot + O_2 \rightarrow ROO\cdot \qquad (6)$$

These radicals can then modify the surface (chain scissions, etching), leading to the ejection of species (ROOH, ROH, R·, ROO·) corresponding to the fragments evidenced on the Al foil, but also potential CO2 and H2O molecules, nevertheless not observed during our plasma treatment [64].

### 4.4. PTFE surface/He post-discharge

**4.4.1. Macroscopic scale**

The XPS spectra indicate no relevant change in the chemical composition of the PTFE surface exposed to the helium post-discharge, and especially no significant defluorination or oxidation [35]. As no oxygenated functions are evidenced, the slight decrease from 114° (native) to 102° (plasma-treated surface) can only be attributed to the smoothening of the surface (i.e. a decrease in the roughness, as represented in figure 10). This assumption is confirmed by AFM measurements which indicate a value of 32.2 nm for the native surface and 27.7 nm after treatment. As already mentioned in the two last sections, the roughness modifications are induced by a surface etching which can induce variable mass losses; very small in this case since it is lower than 1 mg. These weight measurements are consistent with the chemical composition of the aluminum foil indicating a $CF_2$ relative composition of 76.8% when the PTFE was treated (versus 0% for the native case). This elevated composition in $CF_2$ does not mean that an elevated mass of polymer fragments has been ejected: it only describes the relative composition of the foil surface.

**4.4.2. Microscopic scale**

As the energy of the VUV photons is close to 20 eV, the C–C and C–F bonds are expected to be broken indifferently and the amount of the atoms ejected from the sample is considered to be the same as in the case of polyethylene (case 1). As the carbon skeletons are recovered by F atoms, the VUV photons mostly interact with these F atoms. This VUV radiation penetrates into the subsurface but cannot cause as much damage as the oxygen radicals do in the case of a He–O2 post-discharge, since their momentum is almost negligible by comparison with that of an O radical.

The excitation potentials of He metastable species—close to 20 eV—are higher than HDPE and PTFE ionization potentials which are about 11 eV and 7 eV, respectively, and much higher than the C–C, C–F and C–H bond energies in PTFE and PE (on the order of 3 to 5 eV) [65]. Westerdahl et al already evidenced the role played by the VUV radiation emitted from a helium microwave plasma on the decrease in the WCA of FEP surfaces (from 103° to 83°) [66]. As it has already been shown that the radiation at 121.5 nm (10.2 eV) had enough energy to induce PTFE surface modifications, the assumption that at atmospheric pressure the first continuum of He could participate into surface processes (main-chain scission, crosslinking, desaturation) is supported.

When a VUV photon is absorbed by a molecule, its electron density can be strongly disrupted and drive to the formation of antibonding molecular orbitals. A C–C bond exposed to this VUV radiation may present this antibonding character and dissociates, leading subsequently to the ejection of $CF_2$ fragments on the aluminum foil. The electronic states of C–F could also present an antibonding character but to a higher energy. As a consequence, the dissociation of C–F would be less likely to occur. To sum up, PTFE can be regarded as a network whose weak points are the C–C bonds (3.6 eV) and the strong points the C–F bonds (4.5–5.3 eV). When a VUV radiation interacts with the fluorine atoms, the absorbed radiative energy leads to the breaking of the weakest bonds. The He metastable species can also participate to the dissociation of C–C and C–F bonds by transferring their potential energy. The absence of oxygenated function on the PTFE surface could be related to the existence of the atomic fluorine sheath protecting the carbon skeleton: steric effects may then occur and explain that no oxygenated function could be detected as in the case of HDPE [67].

### 4.5. PTFE surface/He–$O_2$ post-discharge

**4.5.1. Macroscopic scale**

The chemical composition of a PTFE surface treated by a He–$O_2$ post-discharge (with a $O_2$ flow rate fixed at 200 mL.min$^{-1}$) does not differ from the case of a pure He post-discharge: no oxygenated function is grafted on the surface. However, the WCA measurements indicate a strong increase from 114° (native) to 150° (plasma-treated), which can only be assigned to an increase in the surface roughness. This assumption is indeed validated by AFM measurements which show a roughness value increasing from 32.2 nm (native surface) to 38.2 nm (plasma-treated surface). The etching of the surface is accompanied by negligible mass losses (even smaller than in the case of a pure He post-discharge). Furthermore, the chemical surface composition of the aluminum foil shows a $CF_2$ content three times lower than in the pure helium treatment, thus confirming the lower mass losses. The etching profile of the treated PTFE surface is represented in figure 10(d).

**4.5.2. Microscopic scale**

No sputtering of the surface is taking place since we did not evidence the ejection of polymer fragments and any mass loss. The atomic structure of the PTFE does not seem damaged by the interactions with the O radicals, $O_2^M$ and $O_2^+$ species because (i) the C–F bonds present BDE much higher than the energy of these incident species, (ii) the F atoms constitute a kind of sheath protecting the carbon backbones from the reactive species of the post-discharge. Such a situation was not encountered in the case of polyethylene, where the O radicals could directly interact with the C–C bonds, thus confirming the assumption of the protective role played by the fluorine atoms. Moreover, this explanation is consistent with the fact that no VUV radiation was detected in the He–$O_2$ post-discharge.

## 5. Conclusions

Polyethylene and various fluoropolymer surfaces have been exposed to a He and a He-$O_2$ post-discharge. O radicals, He metastable atoms and VUV photons are mainly produced in the first case while O radicals, $O_2^M$ and $O_2^+$ species are produced in the other case. Those two treatments are responsible for different etching mechanisms evidenced by correlating XPS spectra, dynamic WCA, mass losses and AFM measurements. We showed that the oxidation and the roughening of a HDPE surface are two phenomena able to occur in synergy (atmospheric helium post-discharge) or in competition (atmospheric He–$O_2$ post-discharge) on the hydrophobic/hydrophilic state of the surface. In the case of a PTFE surface exposed to the He–$O_2$ post-discharge, we showed that its oxidation was negligible, making roughening the only possible factor explaining the increase in the surface hydrophobicity. We also evidenced by optical emission spectroscopy the VUV radiation of $He_2^*$ excimers produced in the discharge and by optical absorption spectroscopy the presence of He metastable species in the pure He post-discharge. Those last species in correlation with the first continuum of $He_2*$ are considered as being responsible for the smoothening of the polymer surface, whatever its nature (HDPE, PVF, PTFE, etc), while the surface etching is mainly attributed to the O radicals.

# 6. Acknowledgments

This work was part of the I.A.P./7 (Interuniversitary Attraction Pole) program 'PSI-Physical Chemistry of Plasma Surface Interactions', financially supported by the Belgian Federal Office for Science Policy (BELSPO). The Mons-Brussels collaboration is supported by the European Commission/Région Wallonne FEDER program (Portefeuille 'Revêtements Fonctionnels'). Research in Mons is also supported by the Science Policy Office of the Belgian Federal Governement (PAI 7/05), the OPTI2MAT Excellence Program of Région Wallonne, and FNRS-FRFC.